\documentclass{article}
\usepackage{ijcai21}

\usepackage{times}
\usepackage{soul}
\usepackage{url}
\usepackage[hidelinks]{hyperref}
\usepackage[utf8]{inputenc}
\usepackage[small]{caption}
\usepackage{graphicx}
\usepackage{amsmath}
\usepackage{amsthm}
\usepackage{booktabs}
\usepackage{algorithm}
\usepackage{algorithmic}

\usepackage{amsfonts}
\usepackage{multirow}
\usepackage{subcaption}
\usepackage{array}

\newcommand{\eat}[1]{}

\title{Discovering Collaborative Signals for Next POI Recommendation with Iterative Seq2Graph Augmentation}

\author{
    Yang Li
    \and
    Tong Chen
    \and
    Yadan Luo
    \and
    Hongzhi Yin
    \and
    Zi Huang
    \affiliations
    The University of Queensland
    \emails
    \{yang.li, tong.chen, h.yin1\}@uq.edu.au, lyadanluol@gmail.com, huang@itee.uq.edu.au
}

\begin{document}

\maketitle

\begin{abstract}
 Being an indispensable component in location-based social networks, next point-of-interest (POI) recommendation recommends users unexplored POIs based on their recent visiting histories. However, existing work mainly models check-in data as isolated POI sequences, neglecting the crucial collaborative signals from cross-sequence check-in information. Furthermore, the sparse POI-POI transitions restrict the ability of a model to learn effective sequential patterns for recommendation. In this paper, we propose Sequence-to-Graph (Seq2Graph) augmentation for each POI sequence, allowing collaborative signals to be propagated from correlated POIs belonging to other sequences. We then devise a novel Sequence-to-Graph POI Recommender (SGRec), which jointly learns POI embeddings and infers a user's temporal preferences from the graph-augmented POI sequence. To overcome the sparsity of POI-level interactions, we further infuse category-awareness into SGRec with a multi-task learning scheme that captures the denser category-wise transitions. As such, SGRec makes full use of the collaborative signals for learning expressive POI representations, and also comprehensively uncovers multi-level sequential patterns for user preference modelling. Extensive experiments on two real-world datasets demonstrate the superiority of SGRec against state-of-the-art methods in next POI recommendation.
\end{abstract}

\section{Introduction}
The fast growth of location-based social networks (LBSNs) facilitates the development of point-of-interest (POI) recommender systems, which help users explore attractive places by capturing their preferences from visiting history \cite{ChengYLK13,FengLZCCY15,YinC16,ChangJKK20}. Among a variety of POI recommendation tasks, next POI recommendation is arguably the most prominent one. Unlike conventional POI recommendation that only considers users' general long-term preferences, next POI recommendation suggests a user the most suitable destinations based on a short check-in trajectory, thus requiring the model to effectively capture dynamic user preferences and account for temporal influences.

\begin{figure}[t]
\centering
\includegraphics[width=\columnwidth]{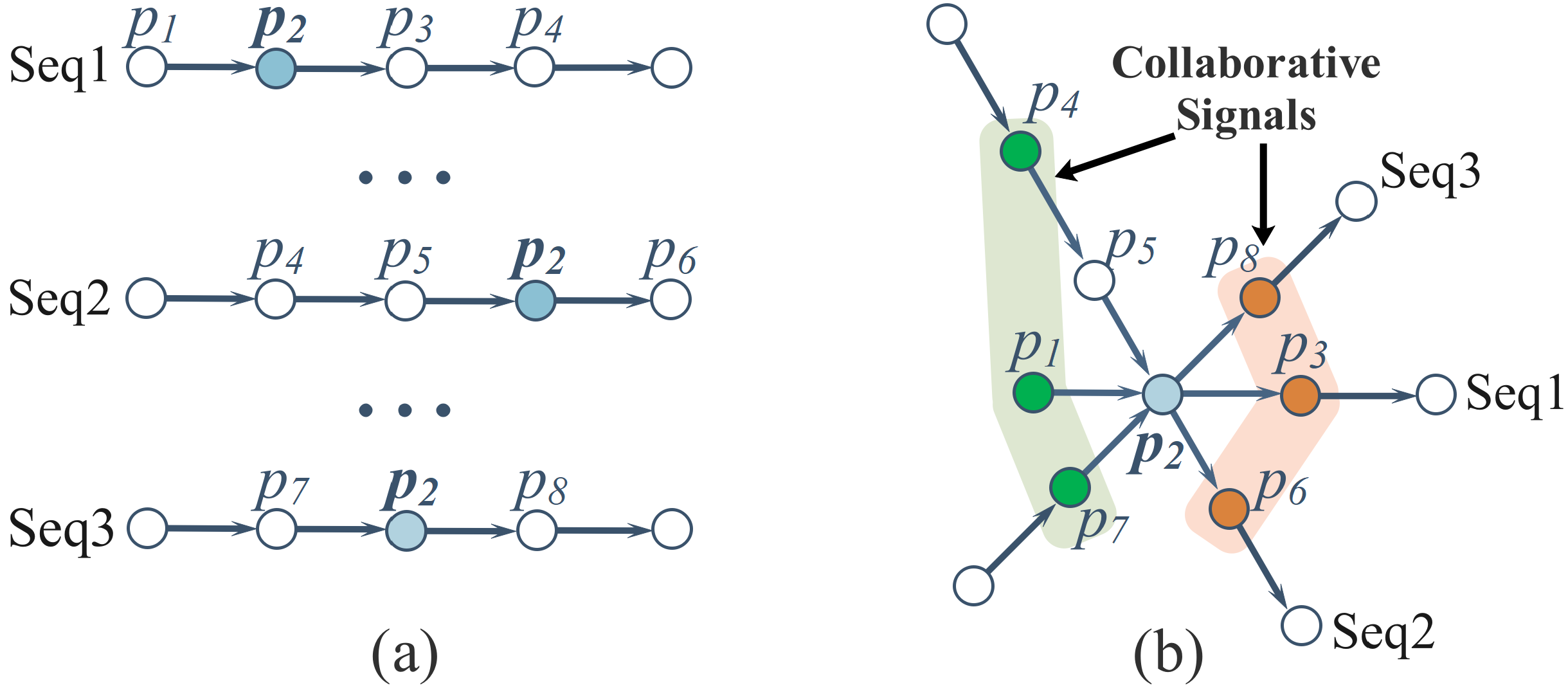}
\caption{(a) Most sequential methods randomly load POI sequences, which neglect the rich contexts from correlated POIs across sequences. (b) In the proposed Seq2Graph augmentation, the collaborative signals among correlated sequences are exploited.}
\label{fig:checkin_seq_example}
\end{figure}

The majority of next POI recommendation approaches are based on sequential models, ranging from the simplistic Markov chain (MC) \cite{ChengYLK13} and SkipGram methods \cite{feng2017poi2vec} to the recently dominating recurrent neural networks (RNNs) \cite{LiSZ18,Zhao0Y0Z0X20}. In a nutshell, those sequential models take an arbitrary user's POI check-in sequence as their input, and learn the latent representations of both POIs and user preferences to facilitate inference of the next POI.

With the demonstrated capability of learning informative POI and user representations from sequential data, RNN-based models are currently the state-of-the-art solutions to most next POI recommendation tasks \cite{LiuWWT16,SunQCLNY20,Zhao0Y0Z0X20,WangYCHWZH20,LiLZSC19}. Despite the versatility of sequential POI recommenders, most existing approaches lack the ability to uncover the \textbf{collaborative signals} \cite{wang2019neural} among relevant POIs when learning POI representations. More concretely, the recommendation process of most sequential approaches is purely conditioned on the POIs within the given sequence, neglecting the rich contexts from highly correlated POIs across sequences. Take Figure \ref{fig:checkin_seq_example}(a) as an example, POIs $p_1$, $p_4$ and $p_7$ from three POI sequences all transit to POI $p_2$, demonstrating strong semantic correlations that can potentially enrich the expressiveness of $p_2$'s learned representation. However, as each POI sequence is assumed to be independent of other sequences in those sequential methods, such high-order collaborative signals (see Figure \ref{fig:checkin_seq_example}(b)) could hardly be captured from all scattered POI sequences, which results in the degradation of model performance. Consequently, given the commonly sparse data on POI-POI transitions \cite{YinZCWZH16,QianLNY19}, the expressiveness of learned POI representation is heavily constrained, impeding the eventual recommendation effectiveness.

Therefore, we aim to address the aforementioned limitations in the existing sequence modelling scheme of next POI recommenders. Instead of straightforwardly modelling independent POI sequences, we innovatively present a solution with graph-augmented POI sequences. Essentially, in a graph-augmented POI sequence, POIs are nodes that are chronologically ordered and associated with neighbour POI nodes via semantic edges, e.g., the one-hop transition relationships observed from other sequences. We refer to this notion as \textbf{Seq2Graph} augmentation in our paper. On the one hand, by bringing the notion of graph-structured data to POI sequences, we can effectively infuse various contexts (e.g., correlated POIs and side information like category tags) into each learned POI representation. As illustrated in Figure \ref{fig:checkin_seq_example}(b), by augmenting the three sequences with Seq2Graph, $p_2$'s surrounding POI nodes offer rich knowledge for describing the properties of $p_2$, hence, the collaborative signals from semantically correlated POIs are thoroughly utilised for learning high-quality POI representations. On the other hand, the capability of learning complex sequential dependencies among POIs is fully retained by the augmented sequences.

To this end, we propose the Sequence-to-Graph POI Recommender (SGRec), a novel solution to next POI recommendation that simultaneously leverages the information from both the graph and sequence sides. In SGRec, we first build a category-aware graph attention layer, which embeds every POI in the sequence by merging the contexts from its neighbour POI nodes. Though in general sequential recommendation, recommenders based on graph neural networks (GNNs) \cite{WuT0WXT19,QiuYHC20,QiuHLY20} start to emerge, the embedding of each item (i.e., node) is mostly learned by aggregating the latent features of co-occurring items from the same input sequence, where the collaborative signals are overlooked. In contrast, SGRec makes full use of the information from neighbour POI nodes sampled from all other POI sequences to learn POI embeddings, which subsequently benefits the learning of user temporal preferences via a position-aware attention mechanism. To further alleviate the data sparsity, we introduce category-awareness into our proposed model. Specifically, we innovatively design an auxiliary objective of learning category-wise transitions. Intuitively, we bind the directional category-category relations with the edges between two connected POIs when learning POI embeddings, and ask the model to predict the category of the upcoming POI, which is in parallel with the next POI prediction task. By gaining knowledge from the much denser category-wise interactions, SGRec can effectively enhance its capability of learning the fine-grained sequential dependencies and users' temporal preferences, thus yielding better performance in next POI recommendation.

The contributions of our paper are three-fold:
\begin{itemize}
    \item We introduce a new take on modelling POI sequences for recommendation by constructing graph-augmented POI sequences to fully capture the collaborative signals from semantically correlated POIs while mining sequential properties. 
    \item We propose SGRec, a novel next POI recommender that selectively aggregates information from both neighbour POI nodes and edges to expressively embed POIs in a sequence. Besides, with an additional focus on learning category-wise transition patterns, SGRec further complements its capability of learning sequential dependencies from sparse POI-level interactions.
    \item We extensively evaluate SGRec on two benchmark datasets, where the results suggest that it outperforms the state-of-the-art baselines with significant margins.
\end{itemize}

\section{Related Work} 
Next POI recommendation mainly explores user-POI interactions and the temporal influences to make recommendation on user's next move. The consecutive check-ins that happen in a relatively short time period carry strong sequential information. \citeauthor{ChengYLK13}[\citeyear{ChengYLK13}] combine first-order Markov chain and matrix factorisation techniques to model POI-POI transitions and user-POI interactions. Recently, recurrent neural networks (RNNs) have demonstrated superior power in sequential data modelling, which drives the studies on exploring user transition patterns. \citeauthor{LiuWWT16}[\citeyear{LiuWWT16}] extend the vanilla RNN with time-specific and distance-specific transition matrices for spatial and temporal pattern modelling. \citeauthor{LiSZ18}[\citeyear{LiSZ18}] use LSTM with attention mechanism to adaptively choose important contextual factors for recommendation. Meanwhile, \citeauthor{ZhaoZLXLZSZ19}[\citeyear{ZhaoZLXLZSZ19}] add time and distance gates into LSTM to capture users' dynamic preferences. 

Different from those existing work that mainly adapts RNNs for sequential pattern mining, which do not explicitly consider the POI correlations across sequences, our work transforms sequences into graph-structured data, thus enabling the proposed GNN-based model to explore the important cross-sequence collaborative information.

\section{Preliminaries}

Let $\mathcal{P}=\{p_1, p_2, ..., p_{|\mathcal{P}|}\}$, $\mathcal{U}=\{u_1, u_2, ..., u_{|\mathcal{U}|}\}$, and $\mathcal{C}=\{c_1, c_2, ..., c_{|\mathcal{C}|}\}$ be the sets of POIs, users and POI categories. We use boldface letters for all latent vectors, e.g., $\textbf{p}$, $\textbf{u}$, $\textbf{c} \in \mathbb{R}^{D}$ to denote the features of the POI $p \in \mathcal{P}$, the user $u \in \mathcal{U}$ and category $c \in \mathcal{C}$, where $D$ is the dimension size.\\
\textit{Definition 1: POI.} A POI represents a specific location, e.g., a cinema. Each POI $p\in\mathcal{P}$ is associated with a category label $c_p \in \mathcal{C}$.\\
\textit{Definition 2: Check-in.} A check-in record can be represented by $s_i = (u, p, c_p, t)$, where $u \in \mathcal{U}$, $p \in \mathcal{P}$, $c_p \in \mathcal{C}$ and $t$ denotes the timestamp when the check-in is observed.\\
\textit{Definition 3: Check-in Sequence.} A check-in sequence is a set of chronologically ordered check-in records of a certain user within a short time interval, e.g., 24 hours. We use $S=\{s_1,s_2,...,s_n\}$ to denote a check-in sequence, where the last check-in $s_n$ is recorded from user's current location $p_n$.\\
\textbf{Problem 1: Next POI Recommendation.} Given a check-in sequence $S_u$ of user $u$, our goal is to recommend $K$ POIs that $u$ is likely to visit at her/his next move.

\begin{figure*}[t]
	\centering
	\includegraphics[width=0.85\textwidth]{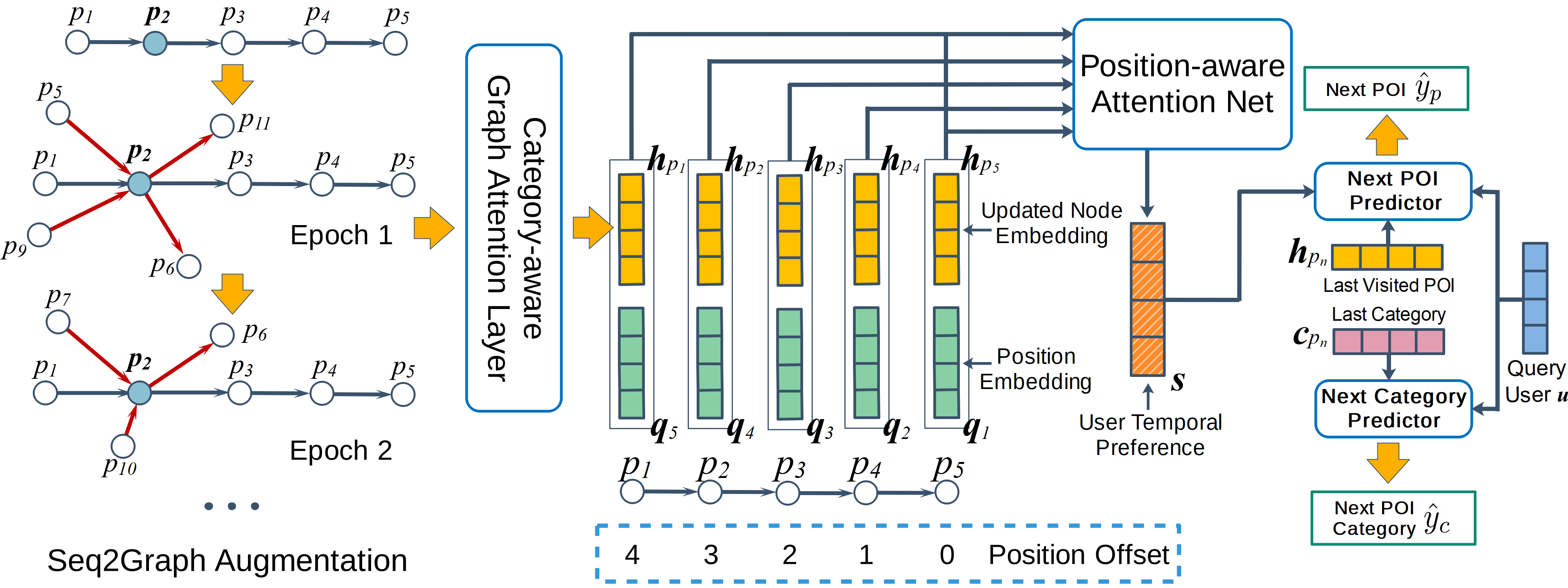}
	\caption{An illustration of our proposed SGRec. Here, we only demonstrate the Seq2Graph augmentation on the node $p_2$ for simplicity. In practice, each node in the sequence will be connected with various correlated neighbours in each training epoch.}
	\label{fig:model_overview}
\end{figure*}

\section{Methodology}
In this section, we first describe the process of Seq2Graph augmentation. Then, we present the details of our proposed SGRec model illustrated in Figure \ref{fig:model_overview}.

\subsection{Seq2Graph Augmentation}
\textbf{Graph-Augmented POI Sequences.} Given a check-in POI sequence $S_u$, each visited unique POI is treated as a node $v_p$, indexed by $p$. Two consecutive POI nodes are connected by a directional edge, i.e., $v_{p_i}\rightarrow v_{p_{i+1}}$ indicating the user moved from the POI $p_i$ to $p_{i+1}$, where $1\leq i\leq n-1$. As shown in Figure \ref{fig:model_overview}, in a graph-augmented POI sequence, each POI node $v_p$ from sequence $S_u$ owns a set of first-order neighbour nodes denoted by $\mathcal{N}(v_p)$. In particular, for $q\in \mathcal{P}$, we have $v_{q}\in \mathcal{N}(v_p)$ if a directional edge $v_{q}\rightarrow v_{p}$ appears in existing POI sequences belonging to the user $u$. Note that $v_p$ is also added into $\mathcal{N}(v_p)$. An augmented sequence of $S_u$ is denoted by $\mathcal{G}_s=\{\mathcal{V}_s, \mathcal{E}_s\} =\{\mathcal{N}(v_{p_1}) \cup \mathcal{N}(v_{p_2}) \cup ... \cup \mathcal{N}(v_{p_n}), \mathcal{E}(v_{p_1}) \cup \mathcal{E}(v_{p_2}) \cup ... \cup \mathcal{E}(v_{p_n})\}$, where we use $\mathcal{V}_s$ and $\mathcal{E}_s$ to represent the sets of all nodes and edges within $\mathcal{G}_s$, and
$\mathcal{N}(v_{p_i})$ and $\mathcal{E}(v_{p_i})$ to denote the sets of all neighbour nodes and edges regarding the $i$-th POI $v_{p_i}$ in the sequence $S_u$, respectively.

\noindent\textbf{Iterative Seq2Graph Augmentation.} 
Though a check-in POI sequence is usually short, each POI in the sequence may appear in multiple POI sequences. Thus, to incorporate richer contextual information and alleviate the data sparsity, we propose to collaboratively investigate different visited POIs with graph-augmented POI sequences. Due to the potentially high computational cost and excessive noise incurred by the sheer amount of neighbour nodes for each POI in the sequence, we present a dynamic neighbour node sampling in SGRec, which obtains a subset of nodes for subsequent computation in each training epoch. Concretely, as shown in the left side of Figure \ref{fig:model_overview}, given one sequence $S_u$, we first obtain all neighbour nodes associated with the original sequence $S_u$, then generate $\mathcal{G}_s$ by uniformly sampling a certain proportion $\gamma$ of nodes from the neighbour sets.
The sampling strategy in Seq2Graph augmentation will be re-executed in each training epoch. As a result, different variants of the graph-augmented POI sequences are generated, bringing diverse cross-sequence contextual information into the given check-in sequence. The augmented sequences are then fed into our proposed SGRec model for learning POI representations and user preferences.

\subsection{Category-Aware Graph Attention Layer}
Introducing category-awareness to GNN could enable the model to learn from the much denser sequential dependencies among POI categories on top of modelling the sparse POI transitions. In SGRec, we design a category-aware attention mechanism when aggregating node information, which seamlessly infuses categorical side information into POI embeddings.
First, we create an embedding vector $\textbf{r}_{c_i \rightarrow c_j} \in \mathbb{R}^{D}$ for each unique category-wise transition relation $c_i \rightarrow c_j$. Instead of defining a separate category sequence, for each POI node $v_p \in \mathcal{G}$, we concatenate the POI embedding with its corresponding category embedding to build a unified node embedding: 
\begin{equation}
	\textbf{v}_p = [\textbf{p};\textbf{c}_p] \in \mathbb{R}^{2D},
\end{equation}
where $\textbf{p}$ and $\textbf{c}_p$ are respectively node $v_p$'s POI and category latent vectors, and $[\cdot;\cdot]$ denotes a concatenation operation. 

One major advantage of SGRec against existing GNN-based sequential recommenders \cite{WuT0WXT19,qiu2019rethinking} is its capability of considering the heterogeneous edge information, i.e, the association between POI categories. Specifically, each directed edge from $\mathcal{E}_s$ corresponds to a pair of categories (e.g., food $\rightarrow$ hotel). As proved in the work of GAT \cite{VelickovicCCRLB18}, selectively aggregating information from the target node's neighbours is beneficial for learning high-quality embeddings. However, it is devised for the nodes with only homogeneous pairwise relationships. Inevitably, in our graph-augmented POI sequences where diverse edges (i.e., different category-wise relations) exist between two POI nodes, quantifying the pairwise attention between nodes will sacrifice such crucial contextual information. 
Thus, we propose a novel attention network, where the interactions between two nodes are additionally conditioned on their edge's properties. In our case, for each POI node $v_p \in \mathcal{G}$ and its neighbour $v_{q}$, the feature of their directional edge $q \rightarrow p$ corresponds to the relation embedding $\mathbf{r}_{c_q \rightarrow c_p}$. We first inject the edge feature into the neighbour node embedding via a non-linear transformation:
\begin{equation}
    \widetilde{\mathbf{v}}_{q} = \textbf{W}_{a}[\textbf{v}_q + \text{MLP}([\textbf{v}_q;\textbf{r}_{c_q \rightarrow c_p}])],
\end{equation}
where $\textbf{w}_{a} \in \mathbb{R}^{2D \times 2D}$ is a trainable weight matrix and $\text{MLP}(\cdot): 4D \rightarrow 2D$ denotes a multi-layer perceptron.

Then, we devise a scoring function $a(\cdot,\cdot)$ to measure the importance of neighbour node $v_{q}$ to the target node $v_p$:
\begin{equation}
    a(v_p, v_{q}) = \mathbf{W}_{gat}^{\top}[\mathbf{W}_{b}\textbf{v}_p;\widetilde{\mathbf{v}}_{q}] + b_{gat},
\end{equation}
where $\mathbf{W}_b \in \mathbb{R}^{2D \times 2D}$ is a transformation matrix, and $\mathbf{W}_{gat} \in \mathbb{R}^{4D}$ and $b_{gat} \in \mathbb{R}$ are the weight matrix and bias.

Afterwards, we use softmax to normalise the calculated attention scores across all $v_p$'s neighbours:
\begin{equation}
    \alpha_{q}=\frac{\exp(\text{LeakyReLU}(a(v_p, v_{q})))}{\sum_{v_m \in \mathcal{N}(v_p)} \exp(\text{LeakyReLU}(a(v_p, v_{m})))}.
\end{equation}
Finally, we obtain the updated node embedding of $v_p$, i.e., $\mathbf{h}_p$, via the following:
\begin{equation}
\label{eq:aggregation}
    \textbf{h}_{p} = \sum_{v_{q} \in \mathcal{N}(v_p)} \alpha_{q} \Phi \textbf{v}_{q},
\end{equation}
where $\Phi \in \mathbb{R}^{2D\times 2D}$ is a trainable weight matrix. For every $v_p \in \mathcal{G}_s$, the produced compact embedding $\textbf{h}_{p} \in \mathbb{R}^{2D}$ encodes the information from both corresponding neighbour POIs and category labels, while the impact from category-wise dependencies and spatial factor is further taken into account during the information aggregation stage in Eq. (\ref{eq:aggregation}). Recall that $\mathbf{h}_p$ is an updated version of $\mathbf{v}_p$, with its first and second halves correspond to embeddings of the POI $p$ (i.e., $\mathbf{p}$) and its category (i.e., $\mathbf{c}_p$). In order to capture higher-order information, we update $\mathbf{p}$ and $\mathbf{c}_p$ using the first and second half of $\mathbf{h}_p$ after every training epoch, respectively:
\begin{equation}
	\mathbf{p} \leftarrow \mathbf{h}_p(1:D),\;\;\;\; \mathbf{c}_p \leftarrow \mathbf{h}_p(D+1:2D),
	\label{eq:iterative_emb}
\end{equation}
where we use $\mathbf{h}_p(a:b)$ to denote the operation of obtaining the values from $a$-th dimension to $b$-th dimension of the vector $\mathbf{h}_p$ and $\leftarrow$ is an assignment operator.
In this way, the topological information of the global POI transition patterns is preserved in the embedding layer.

\subsection{User Temporal Preference Encoding}
So far, the latent POI representation $\textbf{h}_{p}$ carries relevant spatial and categorical perspectives. However, in POI recommendation, the final decision is also heavily associated with users' short-term preferences. \cite{WuT0WXT19} applies an attention mechanism that estimates the impact of historical data points to the last one. However, simply applying the attention mechanism on all POI embeddings hardly capture sufficient sequential and spatial properties from the data. To this end, inspired by \cite{ZhangZCAM17}, we design a position-aware attention mechanism that encodes the sequential information into the final embedding via positional embeddings. Intuitively, the impact of previous check-ins on the current one will degrade through time. For example, given two sequences: $p_1 \rightarrow \mathbf{p_2} \rightarrow \mathbf{p_4}$ and $p_1 \rightarrow \mathbf{p_2} \rightarrow p_3 \rightarrow p_5 \rightarrow p_6 \rightarrow \mathbf{p_4}$. $p_2$'s influence to $p_4$ is apparently higher in the first sequence. In this regard, we define a positional embedding matrix $Q=[\mathbf{q}_1,\mathbf{q}_2,...,\mathbf{q}_n] \in \mathbb{R}^{n \times D}$, which assigns a unique latent embedding to each offset position. Here, $n$ equals to the maximum length of the truncated sequences in the dataset. Specifically, for the $i$-th POI $p_i$ in $S_u$, we first assign the position embedding $\mathbf{q}_{n-i+1}$ to it, where $n-i$ is the offset between the $i$-th check-in and the last one. Then, we build an attention scoring function $b(\cdot,\cdot)$ that incorporates the sequential information (i.e., relative position) to estimate the impact of $p_i$ to the last POI $p_n$:
\begin{equation}
	\begin{aligned}
	b(v_{p_i}, v_{p_n}) = \mathbf{W}_{pat}^{\top} \textrm{tanh}(\mathbf{W}_{h1}^{\top} \mathbf{h}_{p_n} &+\mathbf{W}_{h2}^{\top} \mathbf{h}_{p_i}\\
	&+\mathbf{W}_{q}^{\top} \mathbf{q}_{n-i+1}),
	\end{aligned}
\end{equation}
where $\{\mathbf{W}_{h1}, \mathbf{W}_{h2}\} \in \mathbb{R}^{2D \times D}$, $\mathbf{W}_{q} \in \mathbb{R}^{D \times D}$ and $\mathbf{W}_{pat} \in \mathbb{R}^D$ are trainable weight matrices. Finally, we encode the user's short-term preferences into a single sequence-level embedding vector $\mathbf{s} \in \mathbb{R}^{2D}$ through an attentively weighted sum across all POI embeddings within $S_u$:
\begin{equation}
	\mathbf{s} =\sum_{i=1}^{n} \beta_{i} \mathbf{h}_{p_i},\;\;\;\;\;
	\beta_{i} = \frac{\exp(b(v_{p_i}, v_{p_n}))}{\sum_{m=1}^{n} \exp(b(v_{p_m}, v_{p_n}))}.
\end{equation}

\subsection{Next POI and Category Prediction}
Regarding the large amount of POI candidates, the sparsity of observable transitions between POIs sets an obstacle for deep models to fully capture useful POI-wise sequential dependencies for recommendation. In contrast, the category tags associated with POIs are significantly denser in terms of both quantity and pairwise transitions. In light of this, we formulate a multi-task learning scheme in SGRec, which jointly predicts the exact POI and the next POI category for the user's next visit. Specifically, we first estimate the probability that each POI will be visited:
\begin{equation}
    \widehat{\textbf{y}}^{poi} = \textrm{softmax}(\mathbf{W}_{p}([\textbf{h}_{p_n} \circ \textbf{s}; \textbf{u}]) + \mathbf{b}_p),
\end{equation}
where $\mathbf{W}_p \in \mathbb{R}^{2D \times |\mathcal{P}|}$ and $\mathbf{b}_p \in \mathcal{R}^{|\mathcal{P}|}$ are the transformation matrix and bias of a fully-connected layer, $\circ$ denotes the element-wise multiplication which fuses the user's sequence-level temporal preference (i.e., $\textbf{s}$) with her/his instant preference (i.e., the most recent POI representation $\textbf{h}_{p_n}$), $\textbf{u}$ is the user's long-term preference embedding.

Besides, the next POI category can be predicted via:
\begin{equation}
    \widehat{\textbf{y}}^{cat} = \textrm{softmax}(\mathbf{W}_c([\mathbf{c}_{p_n}; \textbf{u}] + \mathbf{b}_c),
\end{equation}
where $\mathbf{W}_c \in \mathbb{R}^{D \times |\mathcal{C}|}$ and $\mathbf{b}_c \in \mathcal{R}^{|\mathcal{C}|}$ are the transformation matrix and bias of a fully-connected layer, and $\mathbf{c}_{p_n} \in \mathbb{R}^{D}$ is the last POI's category embedding. The rationale of designing this auxiliary task is to enhance our model's capability of capturing category-wise transitions, which is a crucial context in next POI recommendation. In the inference phase, we only perform POI prediction, and select $K$ POIs with the highest probabilities in $\widehat{\textbf{y}}^{poi}$ as the top-$K$ recommendation result.

\subsection{Model Optimisation}
We apply the cross-entropy loss function to quantify the error of both POI and category prediction tasks for SGRec:
\begin{equation}
    \mathcal{L} = - \frac{1}{M} \sum_{m=1}^{M} \underbrace{\textbf{y}^{poi\top}_m \log(\widehat{\textbf{y}}^{poi}_m)}_{\text{next poi loss}} + \eta \underbrace{\textbf{y}^{cat\top}_m \log (\widehat{\textbf{y}}^{cat}_m)}_{\text{next category loss}} +\lambda\|\Psi\|^{2},
\end{equation}
where $m\leq M$ is the index of training samples, $\textbf{y}^{poi}_m$ and $\textbf{y}^{cat}_m$ are the one-hot vectors of POI and category ground truths respectively. $\eta$ is a hyperparameter that decides the weight of next category loss, and $\Psi$ is the set of all trainable parameters for $L2$ regularisation under the control of $\lambda$. 
Such multi-task training objective enhances our model's capability of bridging the crucial recommendation context (i.e., the upcoming POI category) with user preferences, thus allowing for accurate POI recommendation under data sparsity.

\section{Experiments}
\subsection{Datasets and Evaluation Metrics}
We evaluate our proposed SGRec on two public real-world check-in datasets, which are originally crawled from Foursquare and Gowalla and have been extensively used in the previous studies. \textbf{Foursquare} dataset\footnote{https://sites.google.com/site/yangdingqi/home} includes check-ins recorded in Tokyo from Apr. 2012 to Feb. 2013. \textbf{Gowalla} dataset\footnote{http://snap.stanford.edu/data/loc-gowalla.html} contains the check-in records over the world from Feb. 2009 to Oct. 2010. We first filter out the inactive users who have less than 10 check-ins and unpopular POIs that are visited by less than 10 users. Then, we sort the check-in records of each user in ascending order of timestamp and split them into sequences when the time interval of two consecutive check-ins is longer than 1 day and 3 days for Gowalla and Foursquare, respectively. We then discard the sequences that have less than 10 check-ins to ensure the sequence length and truncate long check-in sequences by setting maximum sequence length $n=20$. Finally, we randomly choose 80\%, 10\% and 10\% of each user's check-in sequences from both datasets as training, validation and test sets. The statistics of both preprocessed datasets are summarised in Table \ref{tab:dataset_statistics}. 

\begin{table}[!t]
\centering
\small
\setlength\tabcolsep{1.2pt}
\resizebox{\columnwidth}{!}{
\begin{tabular}{lcccccc}
\hline
\multicolumn{1}{c}{Dataset}    & \#User & \#POI  & \#Cat. & \#Check-in & \#Seq. & \#Relation\\ \hline
Foursquare & 2,203  & 2,876  & 181 & 332,077 & 6,860 & 4,018  \\ \hline
Gowalla    & 24,399 & 55,715 & 370 & 1,699,087  & 15,213 & 30,184    \\ \hline
\end{tabular}
}
\caption{Statistics of Foursquare and Gowalla Datasets}
\label{tab:dataset_statistics}
\end{table}

We choose Hit Ratio at Rank $K$ (HR$@$K) and Normalised Discounted Cumulative Gain (nDCG$@$K) at Rank $K$ on top-$K$ ranked POIs, which are commonly adopted in information retrieval and recommender systems \cite{chen2019air,chen2020sequence} for top-$K$ performance and overall ranking performance, respectively. In each test check-in sequence, the last check-in record is used as the ground truth, and the remaining parts are fed into the model for prediction. Note that the Seq2Graph augmentation is not applied in the test.

\subsection{Baseline Methods}
We compare SGRec with the most representative next POI approaches: 1) \textbf{FPMC-LR} \cite{ChengYLK13}: The model combines the first-order Markov chain and matrix factorisation for next POI recommendation. 2) \textbf{GRU} \cite{ChoMGBBSB14}: A popular variant of RNN, which controls the information flow using two gates. 3) \textbf{TMCA} \cite{LiSZ18}: The model uses LSTM equipped with a multi-level attention module to fuse each POI information into a sequence vector. 4) \textbf{ST-CLSTM} \cite{ZhaoZLXLZSZ19}: A state-of-the-art method that introduces two additional gates in LSTM on the time interval and distance to capture the spatial-temporal contexts. 5) \textbf{SR-GNN} \cite{WuT0WXT19}: Currently, there is no existing GNN-based next POI recommenders. Thus, we consider SR-GNN, a GNN-based model for sequential recommendation, as our baseline method.

\begin{table*}[!htb]
\small
\centering
\renewcommand{\arraystretch}{1.1}
\setlength\tabcolsep{2pt}
\resizebox{\textwidth}{!}{
\begin{tabular}{|c|c|c|c|c|c|c|c|c||c|c|c|c|c|c|c|c|}
\hline
\multirow{3}{*}{Method} & \multicolumn{8}{c||}{Foursquare}  & \multicolumn{8}{c|}{Gowalla}\\
\cline{2-17}
& \multicolumn{4}{c|}{HR$@$K} & \multicolumn{4}{c||}{nDCG$@$K} & \multicolumn{4}{c|}{HR$@$K} & \multicolumn{4}{c|}{nDCG$@$K}\\
\cline{2-17}
& K=1 & K=5 & K=10 & K=20 & K=1 & K=5 & K=10 & K=20 & K=1 & K=5& K=10 & K=20 & K=1 & K=5 & K=10 & K=20\\
\hline
FPMC-LR& 0.106 & 0.135 & 0.141  & 0.152  & 0.057 & 0.123 & 0.143  & 0.159  & 0.026 & 0.050 & 0.062  & 0.078  & 0.040 & \underline{0.060} & \underline{0.068}  & 0.072  \\
GRU& 0.120 & 0.240 & 0.287  & 0.345  & 0.120 & 0.181 & 0.194  & 0.207  & 0.038 & 0.069 & 0.087  & 0.109  & 0.038 & 0.054 & 0.059  & 0.064  \\
TMCA& 0.128 & 0.246 & 0.289  & 0.342  & 0.128 & 0.191 & 0.205  & 0.218  & 0.043 & 0.068 & 0.078  & 0.091  & 0.043 & 0.056 & 0.059  & 0.062  \\
ST-CLSTM& 0.167 &\underline{0.337} &\textbf{0.395} &0.415 & 0.167 & \underline{0.258} & \underline{0.276}  & \underline{0.289}  & \underline{0.056} & \underline{0.079} & \underline{0.096}  & \underline{0.115}  & \underline{0.056} & \underline{0.060} & 0.066 & \underline{0.075}  \\
SR-GNN&  \underline{0.175} & 0.316 & 0.370 & \underline{0.427} & \underline{0.175} & 0.250 & 0.268  & 0.276  & 0.041 & 0.066 & 0.079 & 0.094  & 0.041 & 0.054 & 0.058  & 0.062  \\
\hline
 \textbf{SGRec}& \textbf{0.182} & \textbf{0.351} & \underline{0.388}  & \textbf{0.448}  & \textbf{0.182} & \textbf{0.277} & \textbf{0.290}  & \textbf{0.307}  & \textbf{0.067} & \textbf{0.118} & \textbf{0.137}  & \textbf{0.151}  & \textbf{0.067} & \textbf{0.085} & \textbf{0.103}  & \textbf{0.105}  \\
 
 \textbf{S$^2$GRec}& \textbf{0.195} & \textbf{0.362} & \textbf{0.402}  & \textbf{0.465}  & \textbf{0.195} & \textbf{0.299} & \textbf{0.319}  & \textbf{0.325}  & \textbf{0.067} & \textbf{0.118} & \textbf{0.137}  & \textbf{0.151}  & \textbf{0.067} & \textbf{0.085} & \textbf{0.103}  & \textbf{0.105}  \\
\hline
Improvement & 4.00\% & 4.15\% & -1.77\% & 4.92\% & 4.00\% & 7.36\% & 5.07\% & 6.23\% & 19.64\% & 49.37\% & 42.71\% & 31.30\% & 19.64\% & 41.67\% & 51.47\% & 40.00\%\\
\hline
\end{tabular}
}
\caption{Performance comparisons on two datasets. In each column, the best and the second-best results are highlighted in boldface and underlined respectively.}
\label{tab:baseline_performance}
\end{table*}

\subsection{Implementation Details}
SGRec is implemented using PyTorch with Nvidia GTX 1080 Ti. For consistency, we apply the same dimension size $D$ for all embeddings and weight matrices. Specifically, we set the $D$ to 120, neighbour sampling rate $\gamma$ to 0.2, category loss weight $\eta$ to 0.2 and the number of stacked CA-GAT layers to 2 and 3 for Foursquare and Gowalla, respectively. All the trainable parameters in our model are optimised using Adam optimiser with the batch size of 64, learning rate of 0.005 and $L2$ regularisation strength $\lambda$ of $1e-5$. 

\subsection{Analysis on Recommendation Effectiveness}

We summarise the evaluation results of all models on next POI recommendation task with Table \ref{tab:baseline_performance}. From the statistics in the table, we can draw the following observations:\\
\textbf{1)} The results on Foursquare and Gowalla datasets show that our proposed SGRec significantly outperforms all baseline methods on both evaluation metrics. On Gowalla dataset, compared with the second-best approach ST-CLSTM, SGRec gains 4.15\% and 7.36\% improvement on HR$@$5 and nDCG$@$5 respectively. On Gowalla dataset, SGRec provides around 35\% improvement on average over the best competitors, which clearly demonstrates the effectiveness of SGRec. It is worth mentioning that the density of observed user-POI interactions of Gowalla (0.001) is substantially lower than that of Foursquare (0.052), the huge performance improvement on Gowalla dataset further proves the strong capability of our model on handling sparse data.\\
\textbf{2)} RNN-based methods (GRU, TMCA and ST-CLSTM) perform better than MC-based method, FPMC-LR mainly because of their ability to capture long-term user preferences. Among all RNN-based methods, ST-CLSTM receives the best results. This is because TMCA merely measures the importance of different POIs within a single sequence yet neglects the time influence. However, due to the lack of capturing the collaborative signals across sequences, those RNN-based methods result in sub-optimal performance.\\
\textbf{3)} SR-GNN receives comparable results as ST-CLSTM. It proves the capability of GNN in unveiling the implicit POI transitions. However, since SR-GNN ignores the temporal sequential signals when modelling POI transition patterns, it is inferior to our method on both datasets.

\begin{table}[!t]
\small
\setlength\tabcolsep{3.5pt}
\resizebox{\columnwidth}{!}{
\begin{tabular}{|c|c|c|c|c|}
\hline
\multirow{2}{*}{Method} & \multicolumn{2}{c|}{Foursquare} & \multicolumn{2}{c|}{Gowalla}    \\ \cline{2-5}
    & HR$@$20 & nDCG$@$20 & HR$@$20 & nDCG$@$20 \\ \hline
SGRec$_{ns}$ & 0.414 & 0.297 & 0.143 & 0.103 \\
SGRec$_{nc}$ & 0.441 & 0.299 & 0.145 & 0.107 \\
SGRec$_{w/o.cagat}$ & 0.387 & 0.279 & 0.143  & 0.105  \\
SGRec$_{w/o.pemb}$ & 0.435 & 0.302 & 0.139 & 0.104 \\
SGRec$_{w/o.pattn}$ & 0.434 & 0.300 & 0.141 & 0.104 \\\hline
Full Version & 0.448 & 0.307 & 0.151 & 0.105 \\ \hline
\end{tabular}
}
\caption{Performance of SGRec variants.}
\label{tab:ablation_study}
\end{table}

\subsection{Ablation Study}

To verify the contribution of each proposed component in SGRec, we implement several degraded versions of SGRec for ablation study. The results of ablation tests on Foursquare and Gowalla datasets are presented in Table \ref{tab:ablation_study}.

After either disabling Seq2Graph augmentation or removing CA-GAT, \textbf{SGRec}$_{ns}$ and \textbf{SGRec}$_{w/o.cagat}$ both suffer from similar decreases in the performance on both datasets due to the loss of collaborative signals. When the next category prediction objective is removed, \textbf{SGRec$_{nc}$} has a clear performance drop proving that learning category-level transitions is effective for alleviating the POI-level sparsity. By removing the position-aware attention module, \textbf{SGRec}$_{w/o.pattn}$ receives performance degradation, which implies that capturing the sequential information and user temporal preferences is of great importance for the next POI task.
\begin{figure}[t]
 \centering
 \subfloat[][]
 {\includegraphics[width=0.5\columnwidth]{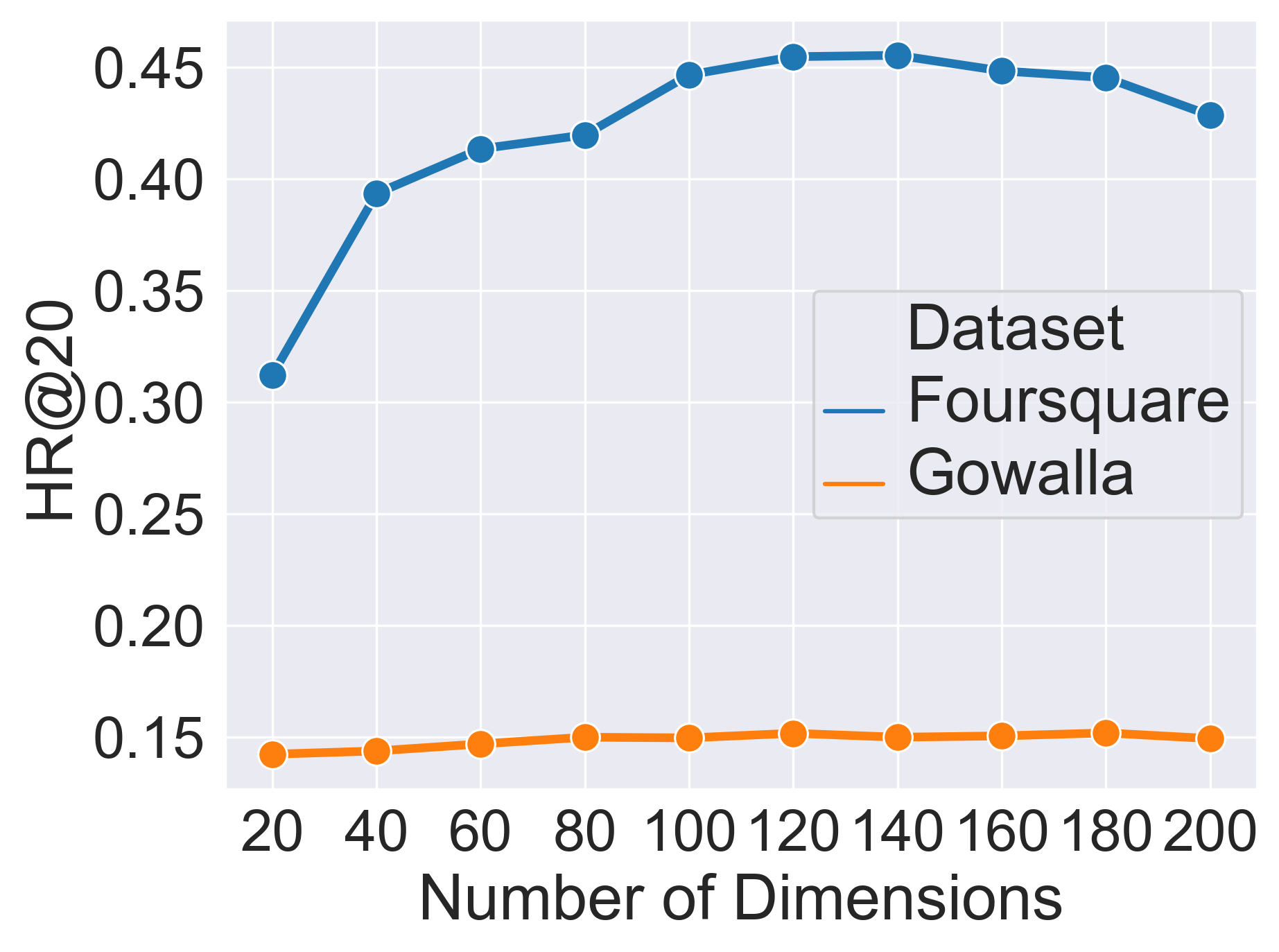}}
 \subfloat[][]
 {\includegraphics[width=0.5\columnwidth]{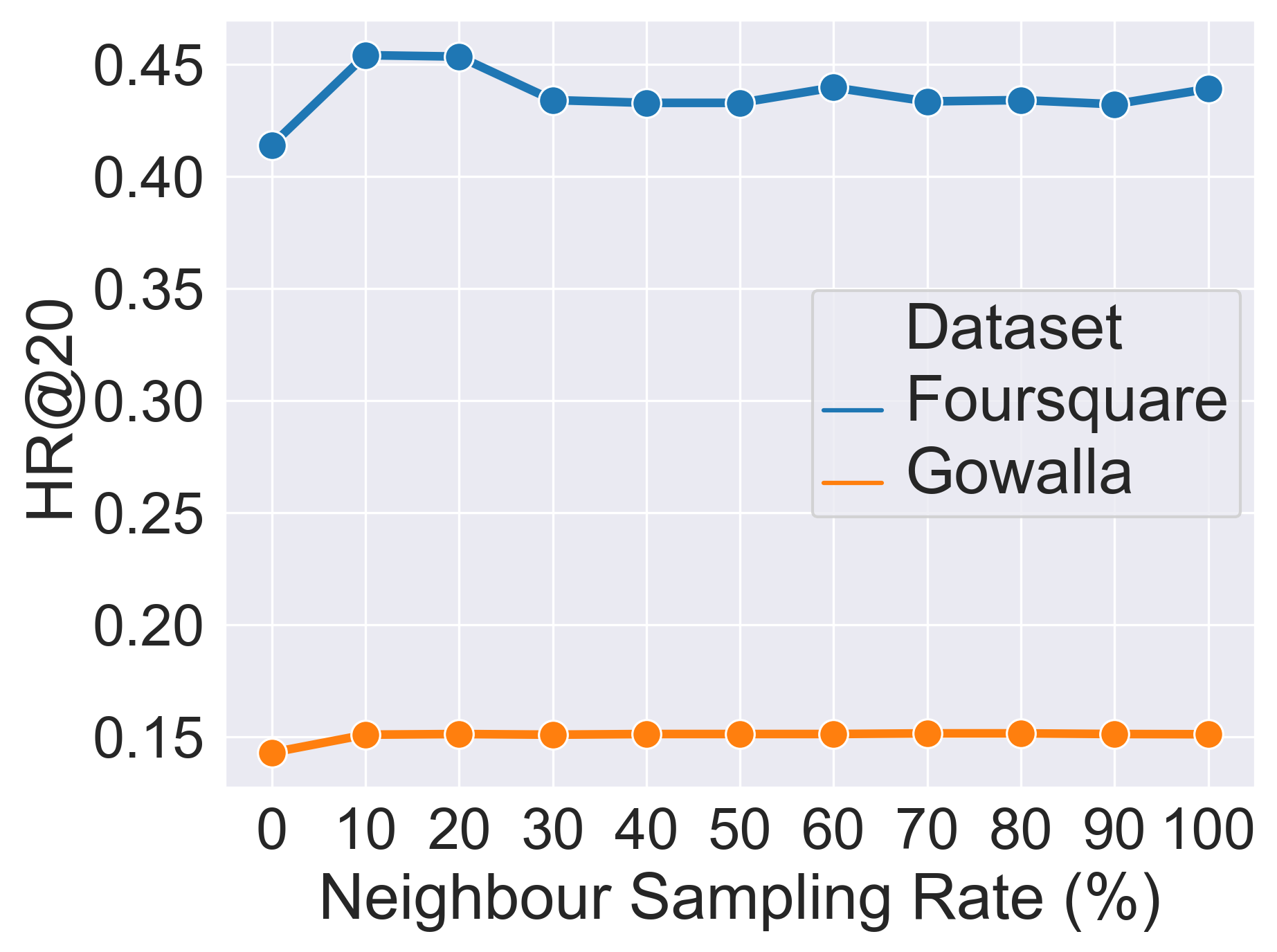}} \\
 \subfloat[][]
 {\includegraphics[width=0.5\columnwidth]{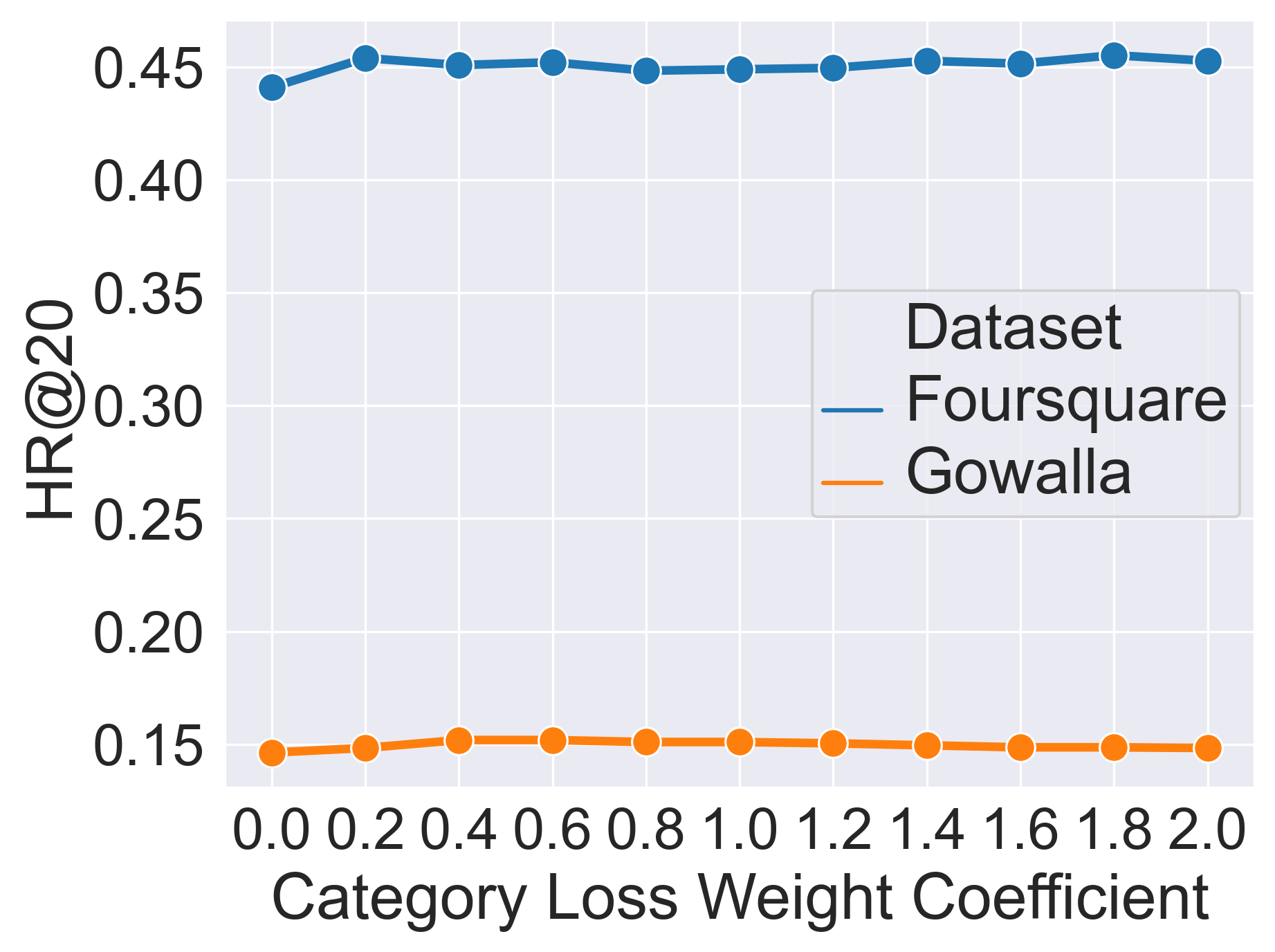}}
 \subfloat[][]
 {\includegraphics[width=0.5\columnwidth]{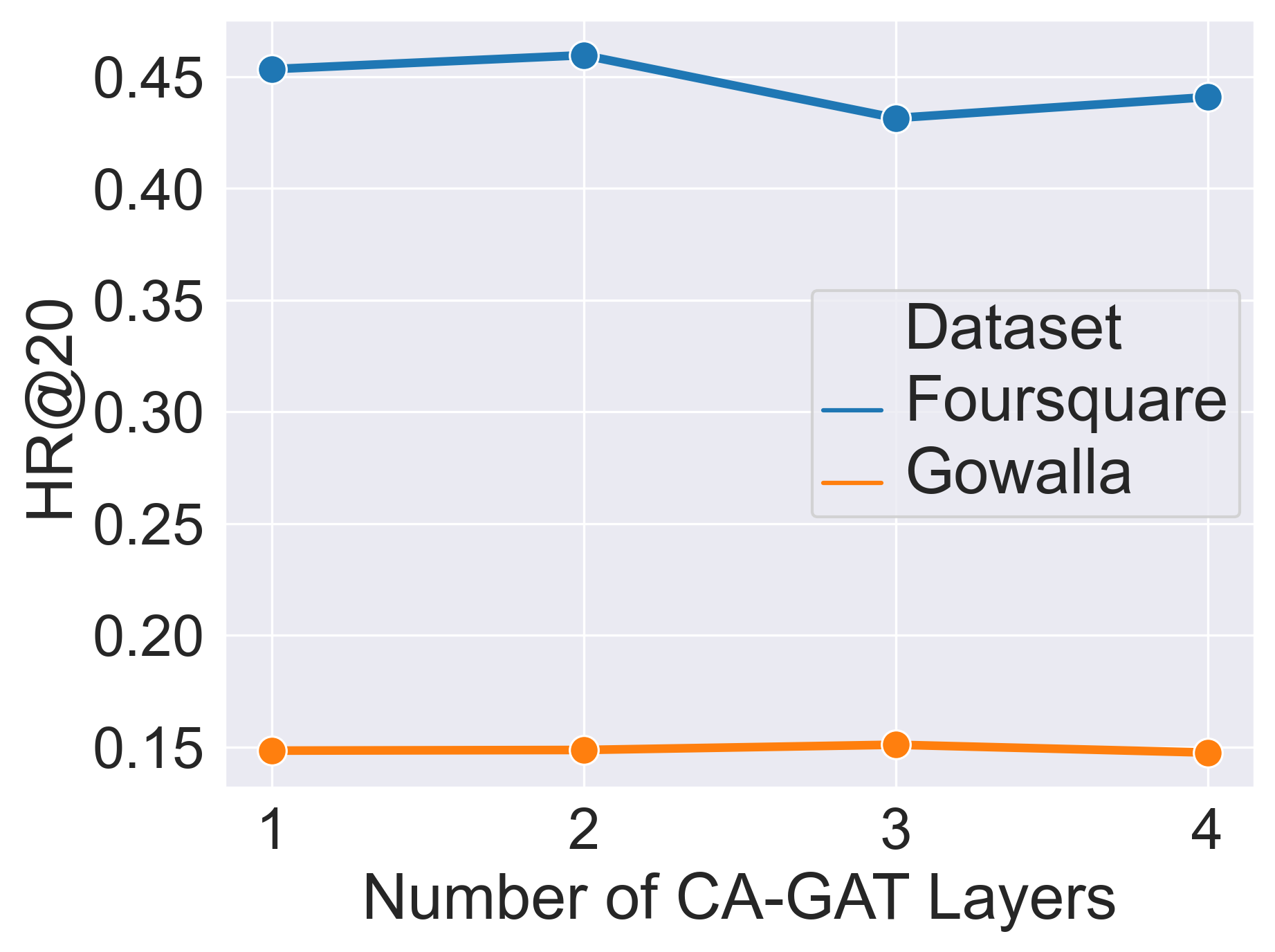}}
 \caption{Impact of (a) dimensionality $D$, (b) neighbour sampling rate $\gamma$, (c) the category loss weight coefficient $\eta$, and (d) the number of stacked CA-GAT layers on Foursquare and Gowalla datasets.}
 \label{fig:hyper_param}
\end{figure}
\subsection{Hyperparameter Analysis}

We further analyse the impact of different hyperparameter values to SGRec, the results are shown in Figure \ref{fig:hyper_param}. Figure \ref{fig:hyper_param}(a) describes HR$@$20 for various dimensionality values ranging from 20 to 200 on two datasets. The model reaches the best performance when $D=120$. We also examine the impact of different neighbour sampling rates. As shown in Figure \ref{fig:hyper_param} (b), the model performance has a clear growth when the rate goes up from 0\% to 10\%. It is worth noting that we disable the iterative embedding update, i.e., Eq.(\ref{eq:iterative_emb}), when $\gamma=0\%$. The result depicts the generalisation capability of SGRec on large-scale datasets since the full power of our model can be exploited with a small number of sampled neighbours. However, the performance has a slight drop when $\gamma \geq 30\%$ on Foursquare. We think this is because sampling too many neighbours will become harmful to preserve the original sequence's information. Figure \ref{fig:hyper_param}(c) shows the impact of the category loss weight coefficient $\eta$, the model has best performance when $\eta=0.4$. Figure\ref{fig:hyper_param}(d) shows that stacking 2 and 3 CA-GAT layers in SGRec can achieve the best performance on both datasets. 
\section{Conclusion}
To deal with the lack of collaborative information and sparse POI-wise interactions within existing sequential POI recommenders, we propose a novel solution named SGRec in this paper. SGRec introduces the notion of Seq2Graph augmentation for incorporating collaborative signals when learning POI embeddings, and leverages its category-awareness to enhance the user preference modelling capacity. Our extensive experiments fully show the effectiveness and robustness of SGRec, demonstrating its strong real-life practicality.
\section*{Acknowledgements}
This work is partially supported by Australian Research Council Discovery Project (ARC DP190102353, DP190101985, CE200100025).

\bibliographystyle{named}
\bibliography{ijcai21}

\end{document}